\documentclass[format=acmsmall, review=false, screen=true]{acmart}

\acmJournal{TOMACS}
\renewcommand\footnotetextcopyrightpermission[1]{} 
\setcopyright{none}
\usepackage{float}
\usepackage{hyperref}
\usepackage{booktabs}
\usepackage{amsmath}
\usepackage{todonotes}
\usepackage{amssymb}
\usepackage[ruled]{algorithm2e}
\usepackage{algorithmic}

\SetAlFnt{\small}
\SetAlCapFnt{\small}
\SetAlCapNameFnt{\small}
\SetAlCapHSkip{0pt}
\usepackage{csquotes}
\usepackage{listings}
\lstdefinestyle{customc}{%
  belowcaptionskip=1\baselineskip,
  breaklines=true,
  xleftmargin=\parindent,
  language=C,
  showstringspaces=false,
  basicstyle=\small\ttfamily,
  keywordstyle=\bfseries\color{green!40!black},
  numberstyle=\tiny,
  commentstyle=\itshape\color{purple!40!black},
  identifierstyle=\bfseries\color{black},
  stringstyle=\color{orange},
   morekeywords={uint64_t,uint32_t,__m256i,__m128i,UINT64_C},
}
\SetAlFnt{\small}
\SetAlCapFnt{\small}
\SetAlCapNameFnt{\small}
\SetAlCapHSkip{0pt}
\IncMargin{-\parindent}
\newcommand{\opcode}[1]{\texttt{#1}}

\usepackage{subfloat,subfig}
\usepackage{tikz}
\usepackage{color}
\usetikzlibrary{chains,positioning,arrows}
\usepackage{siunitx}
\usepackage{url}
\graphicspath{{./gnuplot/}}

\usepackage{etoolbox}

\makeatletter

\begin{document}
\title[Fast Random Integer Generation in an Interval]{Fast Random Integer Generation in an Interval}

\author{Daniel Lemire}
\orcid{0000-0003-3306-6922}
\affiliation{%
  \institution{Universit\'e du Qu\'ebec (TELUQ)}
  \streetaddress{5800 Saint-Denis, Office 1105}
  \city{Montreal}
  \state{Quebec}
  \postcode{H2S 3L5}
  \country{Canada}}
\email{lemire@gmail.com}

\begin{abstract}
In simulations, probabilistic algorithms and statistical tests, we often generate random integers in an interval (e.g., $[0,s)$). For example, random integers in an interval are essential to the Fisher-Yates random shuffle. Consequently, popular languages like Java, Python, C++, Swift and Go include ranged random integer generation functions as part of their runtime libraries.

Pseudo-random values are usually generated in words of a fixed number of bits (e.g., 32~bits, 64~bits) using algorithms such as a linear congruential generator.
We need functions to convert such random words to random integers in an interval ($[0,s)$) without introducing statistical biases. 
The standard functions in programming languages such as Java involve integer divisions.
Unfortunately,  division instructions are relatively expensive. We review an unbiased function to generate ranged integers from a source of random words that avoids
integer divisions with high probability.  To establish the practical usefulness of the approach, we show that this algorithm can multiply the speed of unbiased random shuffling on x64 processors.   
Our proposed approach has been adopted by the Go language for its implementation of the shuffle function.
\end{abstract}

%
%


%
%



\maketitle

\thispagestyle{empty}

\section{Introduction}
\label{sec:intro}
There are many efficient techniques to generate high-quality pseudo-random numbers such as Mersenne Twister~\cite{Matsumoto:1998:MTE:272991.272995}, Xorshift~\cite{marsaglia2003xorshift,Panneton:2005:XRN:1113316.1113319}, 
linear congruential generators~\cite{l1999tables,LEcuyer:1993:SGM:169702.169698,de1988parallelization,fishman2013monte,LEcuyer:1990:RNS:84537.84555}
and so forth~\cite{l2017random,l2012random}. 
Many pseudo-random number generators produce 32-bit or 64-bit words that can be interpreted as integers in $[0,2^{32})$ and $[0,2^{64})$ respectively:
the produced values are practically indistinguishable from truly random numbers in $[0,2^{32})$ or $[0,2^{64})$. In particular, no single value is more likely than any other.

However, we often need random integers selected uniformly from an interval $[0,s)$, and this interval may change dynamically. It is useful for selecting
an element at random in an array containing $s$~elements, but there are less trivial uses. For example, the Fisher-Yates random shuffle described by Knuth~\cite{Knuth1969,Durstenfeld:1964:ARP:364520.364540} (see Algorithm~\ref{algo:knuthshuffle}) requires one random integer in an interval for each value in an array to be shuffled. Ideally, we would want these values to be generated without bias so that all integer values in $[0,s)$ are equally likely. Only then are all permutations equally likely. A related algorithm is \emph{reservoir sampling}~\cite{Vitter:1985:RSR:3147.3165} (see Algorithm~\ref{algo:reservoirsampling}) which randomly selects a subset of values from a possibly very large array, even when the size of the array is initially unknown.

We use random permutations as part of simulation algorithms~\cite{Devroye:1997:RVG:268403.268413,Calvin:1998:UPR:280265.280273,Owen:1998:LSS:272991.273010,Osogami:2009:FPB:1540530.1540533,Amrein:2011:VIS:1899396.1899401,Hernandez:2012:CNO:2379810.2379813}.
The performance of randomized permutation algorithms is important. 
Various non-parametric tests in
statistics and machine learning repeatedly
permute randomly the original data. In some contexts, 
Hinrichs et al.\ found that the computational burden due to random permutations can be prohibitive~\cite{Hinrichs:2013:SUP:2999611.2999711}. 
Unsurprisingly, much work has been done
on parallelizing permutations~\cite{Shterev2010,Langr:2014:A9P:2684421.2669372,SANDERS1998305,Gustedt2008,Waechter2012} for greater speed.

\begin{algorithm}[t]
\begin{algorithmic}[1]
\REQUIRE array $A$ made of $n$ elements indexed from $0$ to $n-1$
\FOR{$i = n-1, \ldots, 1$}
\STATE $j \leftarrow $ random integer in $[0,i]$
\STATE exchange $A[i]$ and $A[j]$
\ENDFOR
\end{algorithmic}
\caption{Fisher-Yates random shuffle\label{algo:knuthshuffle}: it shuffles an array of size $n$ so that $n!$~possible permutations are equiprobable.}
\end{algorithm}

\begin{algorithm}[t]
\begin{algorithmic}[1]
\REQUIRE array $A$ made of $n$ elements indexed from $0$ to $n-1$
\REQUIRE integer $k$ ($0 < k \leq n$)
\STATE $R\leftarrow $ array of size $k$
\FOR{$i = 0, \ldots, k-1$}
\STATE $R[i] \leftarrow A[i]$
\ENDFOR
\FOR{$i = k, \ldots, n-1$}
\STATE $j \leftarrow $ random integer in $[0,i]$
\IF{$j < k$}
\STATE $R[j] \leftarrow A[i]$
\ENDIF
\ENDFOR
\STATE \textbf{return} $R$
\end{algorithmic}
\caption{Reservoir sampling\label{algo:reservoirsampling}: returns an array $R$ containing $k$~distinct elements picked randomly from an array $A$ of size $n$ so that all $n \choose k$~possible samples are equiprobable.}
\end{algorithm}

One might think that going from fixed-bit pseudo-random numbers (e.g., 32-bit integers) to pseudo-random numbers in an interval is a minor, inexpensive operation. However, this may not be true, at least when the interval is a changing parameter and we desire a uniformly-distributed result.
\begin{itemize}
\item Let us consider a common but biased approach to the problem of converting numbers from a large interval $[0,2^n)$ to numbers in a subinterval $[0,s)$ ($s \leq 2^n$): the modulo reduction $x \to x \bmod s$. On x64 processors, this could be implemented through the division (\opcode{div}) instruction when both $s$ and $x$ are parameters. When applied to 32-bit registers, this instruction has a latency of 26~cycles~\cite{fog2016instruction}. With 64-bit registers, the latency ranges from 35 to 88~cycles, with longer running times for small values of $s$. 
\item Another biased but common approach consists in using a fixed-point
floating-point representation consisting of the following step:
\begin{itemize}
\item we convert the random word to a floating-point number in the interval $[0,1)$, 
\item we convert the integer $s$ into a floating-point number,
\item we multiply the two resulting floating-point numbers,
\item and we convert the floating-point result to an integer in $[0,s)$.
\end{itemize}
When using the typical floating-point standard (IEEE 754), we can at best represent all values in $[0,2^{24})$ divided by $2^{24}$ using a 32-bit floating-point number.  Thus we do not get the full 32-bit range: we cannot generate all numbers in $[0,s)$ if $s>2^{24}$. To do so, we must use double precision floating-point numbers, and then we can represent all values in $[0,2^{53})$ divided by $2^{53}$. Moreover converting between floating-point values and integers is not without cost: the corresponding instructions on x64 processors (e.g., \texttt{cvtsi2ss}, \texttt{cvttss2si}) have at least six~cycles of latency on Skylake processors~\cite{fog2016instruction}.
\end{itemize}
While generating a whole new 64-bit pseudo-random number can take as little as a handful of cycles~\cite{Saito2008}, transforming it into an integer in an interval ($[0,s)$ for $s \in [0,2^{64})$) without bias can take an order of magnitude longer when using division operations. 


There is a fast technique that avoids division and does not require floating-point numbers. Indeed, given an integer $x$ in the interval $[0,2^L)$, we have that the integer $( x  \times s ) \div 2^L$ is in $[0,s)$ for any integer $s\in[0,2^L]$. If the integer $x$ is picked at random in $[0,2^L)$, then the result $( x  \times s ) \div 2^L$ is a random integer in $[0,s)$~\cite{overton2011}. The division by a power of two ($\div 2^L$) can be implemented by a bit shift instruction, which is inexpensive.
A multiplication followed by a shift is much more economical on current processors than a division, as it can be completed in only a handful of cycles. 
It introduces a bias, but we can correct for it efficiently using the  rejection method (see \S~\ref{sec:avoiding}).
This multiply-and-shift approach is similar in spirit to the multiplication by a floating-point number in the unit interval ($[0,1)$) in the sense that
$(s x) \div 2^L$  can be intuitively compared with  $s \times x/2^L$ where
$x/2^L$ is a random number in $[0,1)$.  


Though the idea that we can avoid divisions when generating numbers in an interval is not novel, we find that many standard libraries (Java, Go, \ldots) use an approach that incurs at least one integer division per function call. We believe that a better default would be an algorithm that avoids division instructions with high probability. We show experimentally that such an algorithm can provide superior performance.

\section{Mathematical Notation}
\label{sec:math}

We let $\lfloor x \rfloor$ be the largest integer smaller than or equal to $x$,
we let $\lceil x \rceil$ be the smallest integer greater than or equal to $x$.
We let $x \div y$ be the integer division of $x$ by $y$, defined as $\lfloor x / y \rfloor$. We define the remainder of the division of $x$ by $y$ as $x \bmod y$: $x \bmod y \equiv x - (x \div y) y$. 

We are interested in the integers in an interval $[0,2^L)$ where, typically,
$L=32$ or $L=64$. We refer to these integers as $L$-bit integers. 

When we consider the values of $x \bmod s$ as
$x$ goes from $0$ to $2^L$, we get the $2^L$ values 
\begin{align*}\underbrace{\overbrace{0,1, \ldots, s-1}^{\text{$s$ values}}, \overbrace{0, 1, \ldots, s-1}^{\text{$s$ values}}, \ldots, \overbrace{0, 1, \ldots, s-1}^{\text{$s$ values}}}_{\text{$(2^L \div s) s$~values}}, \overbrace{0, 1, \ldots, (2^L \bmod s) - 1}^{\text{$2^L \bmod s$ values}}. 
\end{align*}
We have the following lemma by inspection. 

\begin{lemma}\label{lemma:tech1}
Given integers $a,b, s >0$, there are exactly $(b - a) \div s$~multiples of $s$ in $[a,b)$ whenever $s$ divides $b-a$. More generally, there are exactly $(b - a) \div s$~integers in $[a,b)$ having a given remainder with $s$ whenever $s$ divides $b-a$. 

%
%
%
\end{lemma}


%

A geometric distribution with success probability $p\in [0,1]$ is a discrete distribution taking value $k\in \{1,2,\ldots\}$ with probability $(1-p)^{k-1} p$. The mean of a geometric distribution is $1/p$.

\section{Existing Unbiased Techniques Found in Common Software Libraries}
\label{sec:existing}

Assume that we have a source of uniformly-distributed  $L$-bit random numbers, i.e., integers in $[0,2^L)$. 
From such a source of random numbers, we want to produce a uniformly-distributed random integer $y$ in $[0,s)$ for some integer $s \in [1,2^L]$. That is all integers from the interval are equally likely: $P(y = z) = 1/s$ for any integer $z \in [0,s)$. We then say that the result in \emph{unbiased}. 

If $s$ divides $2^L$, i.e., it is a power of two, then we can  divide the random integer $x$ from $[0,2^L)$ by $2^L/s = 2^L \div s$. However, we are interested in the general case where $s$ may be any integer value in $[0,2^L)$. 

We can achieve an unbiased result by the rejection method~\cite{von1961various}. For example, we could generate random integers $x \in [0,2^L)$ until  $x$ is in $[0,s)$, rejecting all other cases.\footnote{For some applications where streams of random numbers need to be synchronized, the rejection method is not applicable~\cite{asmussen2007stochastic,bratley2011guide,law2007,fu2017history,l2015random}.}  Rejecting so many values is wasteful. Popular software libraries use more efficient algorithms. We provide code samples in Appendix~\ref{appendix:codesamples}.

\subsection{The OpenBSD Algorithm}

The C standard library in OpenBSD and macOS have an \texttt{arc4random\_uniform} function to generate unbiased random integers in an interval $[0,s)$. 
See Algorithm~\ref{algo:arc4}. The Go language (e.g., version 1.9) has adopted the same algorithm for its \texttt{Int63n} and \texttt{Int31n} functions, with minor implementation differences~\cite{gorand}.
The GNU C++ standard library (e.g., version 7.2) also relies on the same algorithm~\cite{gnucpplib}. 

The interval $[2^L \bmod s, 2^L)$ has size $2^L-(2^L \bmod s)$ which is divisible by $s$.
From Lemma~\ref{lemma:tech1}, if we generate random integers from integer values in $[2^L \bmod s, 2^L)$ as remainders of a division by 
$s$ ($x \bmod s$), then each of the integers  occur 
for $2^L \div s$~integers $x\in [0,2^L)$. To produce 
integers in $[2^L \bmod s, 2^L)$, we use the rejection method: 
we generate integers in $x \in [0, 2^L)$ but reject the result 
whenever $x<2^L \bmod s$. If we have a source of unbiased 
random integers in $[0, 2^L)$, then the result of Algorithm~\ref{algo:arc4}
is an unbiased random integer in $[0,s)$. 


The number of random integers consumed by this algorithm follows  a geometric distribution, with a success probability $p=1-(2^L \bmod s)/2^L $. On average, we need $1/p$~random words. This average is less than two, irrespective of the value of $s$.
The algorithm always requires the computation of two remainders.

\begin{algorithm}[t]
\begin{algorithmic}[1]
\REQUIRE source of uniformly-distributed random integers in $[0,2^L)$
\REQUIRE target interval $[0,s)$ with $s \in [0,2^L)$
\STATE $t \leftarrow (2^L - s) \bmod s$\hfill{}\COMMENT{$(2^L - s) \bmod s =  2^L \bmod s $} 
\STATE $x\leftarrow$ random integer in $[0,2^L)$
\WHILE[Application of the rejection method]{$x < t$}
\STATE $x\leftarrow$ random integer in $[0,2^L)$
\ENDWHILE
\STATE \textbf{return} $x \bmod {s}$
\end{algorithmic}
\caption{The OpenBSD algorithm. \label{algo:arc4}}
\end{algorithm}


There is a possible trivial variation on the algorithm where instead
of rejecting the integer from $[0,2^L)$ when it is part of the first $2^L \mod s$~values (in $[0,2^L\bmod s)$), 
we reject the integer from $[0,2^L)$ when it is part of the last $2^L\mod s$~values (in $[2^L - (2^L \mod s),2^L)$). 


%

%
%
%
%
%
%
%

\subsection{The Java Approach}

It is unfortunate that Algorithm~\ref{algo:arc4} always requires the computation 
of two remainders, especially because we anticipate such computations to have high latency. The first remainder is used to determine whether a rejection is necessary ($x< (2^L-s) \bmod s$), and the second 
remainder is used to generate the value in $[0, s)$ as $x \bmod  s$.

The Java language, in its \texttt{Random} class, uses an approach that often requires a single
remainder (e.g., as of OpenJDK~9).
Suppose we pick a number $x\in [0, 2^L)$ and we compute 
its remainder $x \bmod s$.
Having both $x$ and $x \bmod s$, we can determine
whether $x$ is allowable (is in $[0, 2^L-(2^L \bmod s))$) without using
another division. When it is the case, we can then 
return $x \bmod s$ without any additional computation. See Algorithm~\ref{algo:java}.


\begin{algorithm}[t]
\begin{algorithmic}[1]
\REQUIRE source of uniformly-distributed random integers in $[0,2^L)$
\REQUIRE target interval $[0,s)$ with $s \in [0,2^L)$
\STATE $x\leftarrow$ random integer in $[0,2^L)$
\STATE $r\leftarrow x \bmod s$ \hfill{}\COMMENT{$x -r > 2^L - s \Leftrightarrow x \in[0, 2^L-(2^L \bmod s)) $}
\WHILE[Application of the rejection method]{$x -r > 2^L - s$}
\STATE $x\leftarrow$ random integer in $[0,2^L)$
\STATE $r\leftarrow x \bmod s$
\ENDWHILE
\STATE \textbf{return} $r$
\end{algorithmic}
\caption{The Java algorithm. \label{algo:java}}
\end{algorithm}

The number of  random words and remainders used by the Java algorithm follows a geometric distribution, with a success probability $p=1-(2^L \bmod s)/2^L $. Thus, on average, we need $1/p$~random words and remainders. Thus when $s$ is small ($s \ll 2^L$), we need $\approx 1$~random words and remainders. This compares favorably to the 
 OpenBSD algorithm that always requires the computation of two remainders. However, the maximal number of divisions required by the OpenBSD algorithm is two, whereas the Java approach could require infinitely many divisions in the worst case.

\section{Avoiding Division}
\label{sec:avoiding}
Though arbitrary integer divisions are relatively expensive on common
processors, bit shifts are less expensive, often requiring just one cycle.
When working with unsigned integers, a bit shift is equivalent to a division
by a power of two. Thus we can compute $x \div 2^k$ quickly for any power of two $2^k$. Similarly, we can compute the remainder of the division by a power of two as a simple bitwise logical AND: $x \bmod 2^k = x \mathrm{~AND~} (2^k-1)$.

Moreover, common general-purpose processors (e.g., x64 or ARM processors) can efficiently
compute the full result of a multiplication. That is, when multiplying
two 32-bit integers, we can get the full 64-bit result and, in particular,
the most significant 32~bits. Similarly, we can multiply two 64-bit integers and get the full 128-bit result or just the most significant
64~bits when using a 64-bit processor.
Most modern programming languages (C, C++, Java, Swift, Go\ldots) have 
native support for 64-bit integers. Thus it is efficient to get the most
significant 32~bits of the product of two 32-bit integers. 
To
get the most significant 64~bits of the product of two 64-bit integers in C and C++, we can use
either intrinsics (e.g., \texttt{\_\_umulh} when using the Visual Studio compiler)
or the \texttt{\_\_uint128\_t} extension supported by the GNU GCC and LLVM's clang compilers. The Swift language has the \texttt{multipliedFullWidth} function that works with both 32-bit and 64-bit integers. It gets compiled to efficient binary code.

Given an integer $x \in [0,2^L)$, we have that
$(x  \times s) \div 2^L \in [0,s)$. 
By multiplying by $s$, we take integer values in the range $[0, 2^L)$ and map them to multiples of $s$ in $[0, s \times 2^L)$. By dividing by $2^L $, we map all multiples of $s$ in $[0,2^L)$ to 0, all multiples of $s$ in $[2^L,2\times 2^L)$ to one, and so forth. The $(i+1)^\mathrm{th}$~interval is $[i \times 2^L, (i + 1) \times 2^L)$. 
By Lemma~\ref{lemma:tech1}, 
there are exactly  $\lfloor 2^L/s \rfloor $  multiples of $s$ in intervals $[i \times 2^L + (2^L \bmod s), (i + 1) \times 2^L)$ since $s$
divides the size of the interval ($2^L-(2^L \bmod s) $).
Thus if we reject the multiples of $s$ that appear in $[i \times 2^L, i \times 2^L + (2^L \bmod s))$, we get that all intervals have exactly $\lfloor 2^L/s \rfloor $  multiples of $s$.
We can formalize this result as the next lemma.

\begin{lemma}\label{lemma:ouralgo}
Given any integer $s\in [0,2^L)$,  we have that for any integer $y \in [0,s)$, there are exactly $\lfloor 2^L/s \rfloor $  values $x \in [0, 2^L)$ such that 
$(x \times s) \div 2^L =y$ and $(x  \times s) \bmod 2^L \geq 2^L \bmod s$. 
\end{lemma}

Algorithm~\ref{algo:divisionless} is a direct application of Lemma~\ref{lemma:ouralgo}. 
It generates unbiased random integers in  $[0, s)$ for any integer $s\in (0, 2^L)$.


%

\begin{algorithm}[t]
\begin{algorithmic}[1]
\REQUIRE source of uniformly-distributed random integers in $[0,2^L)$
\REQUIRE target interval $[0,s)$ with $s \in [0,2^L)$
\STATE $x\leftarrow$ random integer in $[0,2^L)$
\STATE $m\leftarrow x \times s$  
\STATE $l\leftarrow m \bmod 2^L$  
\IF[Application of the rejection method] {$l <s$}
\STATE $t \leftarrow (2^L - s) \bmod s$\hfill{}\COMMENT{$2^L \bmod s = (2^L-s)\bmod s$}
\WHILE{$l < t$}
\STATE $x\leftarrow$ random integer in $[0,2^L)$
\STATE $m\leftarrow x \times s$  
\STATE $l\leftarrow m \bmod 2^L$  
\ENDWHILE
\ENDIF
\STATE \textbf{return} $m \div 2^L$
\end{algorithmic}
\caption{An efficient algorithm to generate unbiased random integers in an interval. \label{algo:divisionless}}
\end{algorithm}

\begin{figure}\centering
\includegraphics[width=0.49\columnwidth]{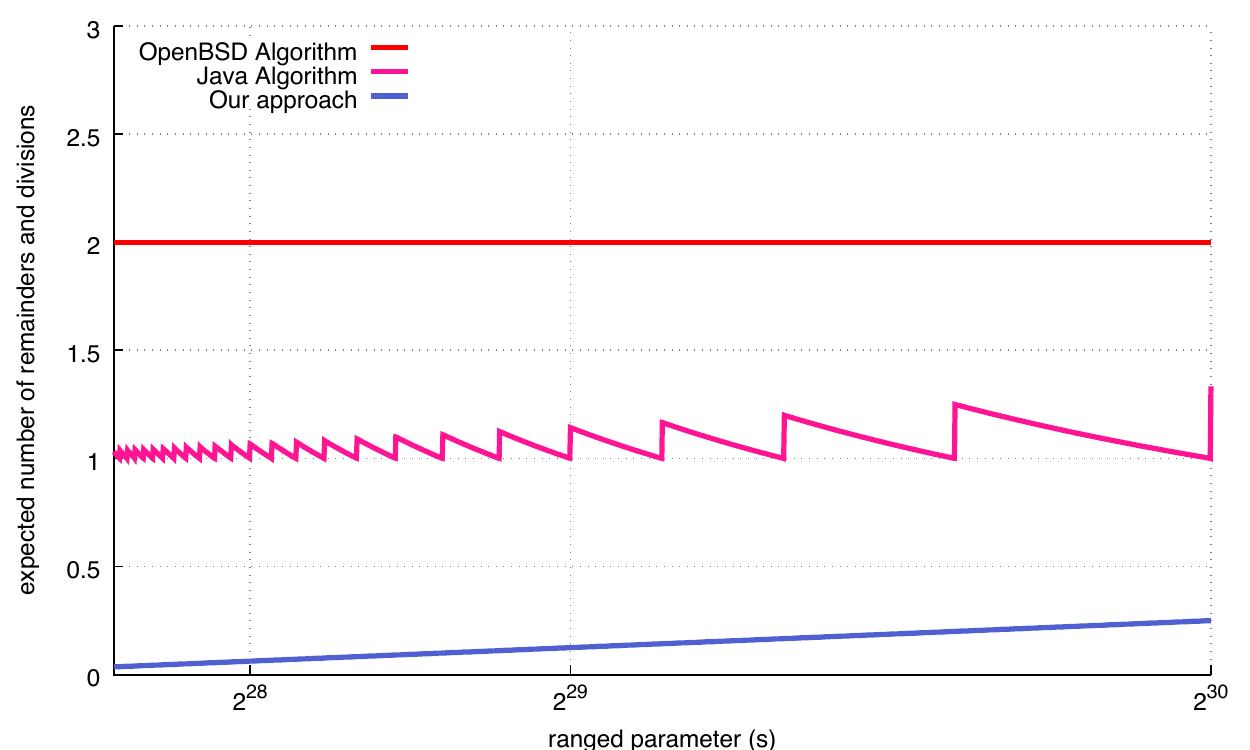}
\caption{\label{fig:expected}Expected number of remainder computations (excluding those by known powers of two) in the production an unbiased random integer in $[0,s)$ for 32-bit integers.}
\end{figure}

This algorithm has the same probabilistic distribution
of random words as the previously presented algorithms, requiring the same number of $L$-bit uniformly-distributed random words on average.
However, the number of divisions (or remainders)
is more advantageous. Excluding divisions by a power of two, we have a probability $\frac{2^L-s}{2^L}$ of using no division at all.
Otherwise, if the initial value of $l=(x  \times s) \bmod 2^L$ is less than $s$, then we incur automatically the cost of a single division (to compute $t$), but no further division (not counting divisions by a power of two).
Table~\ref{table:expected} and Fig.~\ref{fig:expected} compare the three algorithms in terms of the number of remainder computations needed.
 
 \begin{table}
 \caption{\label{table:expected}
 Number of remainder computations (excluding those by known powers of two) in the production an unbiased random integer in $[0,s)$.}
\centering
\begin{tabular}{cccc}\toprule
                   &        \multicolumn{1}{p{2.8cm}}{expected number of remainders  per integer in $[0,s)$ }  &     \multicolumn{1}{p{2.8cm}}{expected number of remainders when the interval is small ($s\ll 2^L$) }    & \multicolumn{1}{p{2.8cm}}{maximal number of remainders per integer in $[0,s)$  } \\\midrule
 OpenBSD  (Algorithm~\ref{algo:arc4})    & 2   & 2  &  2     \\[1.5em] 
 Java      (Algorithm~\ref{algo:java})        &    $\frac{2^L}{2^L-(2^L  \bmod s)}$      & 1 & $\infty$\\[1.5em] 
 Our approach  (Algorithm~\ref {algo:divisionless})     &   $\frac{s}{2^L}$ & 0  & 1  \\
 \bottomrule
 \end{tabular}
 \end{table}

\section{Experiments}

We implemented our software in C++ on a Linux server with an Intel (Skylake) i7-6700 processor running at \SI{3.4}{GHz}.
This  processor has \SI{32}{kB} of L1 data cache, \SI{256}{kB} of L2 cache per core with \SI{8}{MB} of L3 cache, and \SI{32}{GB} of RAM (DDR4~2133, double-channel). We use the GNU GCC~5.4 compilers with  the ``\texttt{-O3 -march=native}'' flags. To ensure reproducibility, we make our software freely available.\footnote{\url{https://github.com/lemire/FastShuffleExperiments}} Though we implement and benchmark a Java-like approach (Algorithm~\ref{algo:java}), all our experiments are conducted using C++ code.

For our experiments, we use a convenient and fast linear congruential generator with the recurrence formula $X_{n+1} = c \times  X_n \bmod 2^{128}$ where $c= 15750249268501108917$ to update the 128-bit state of the generator ($X_i\in [0,2^{128})$ for $i=0, 1, 2, \ldots$), returning $X_{n+1}\div 2^{64}$ as a 64-bit random word~\cite{l1999tables}. We start from  a 128-bit seed $X_0$. This well-established generator passes difficult statistical tests such as Big~Crush~\cite{LEcuyer:2007:TCL:1268776.1268777}. It is well suited to x64 processors because they have fast 64-bit multipliers.

We benchmark the time required, per element, to randomly shuffle arrays of integers having different sizes. We can consider array indexes to be either 32-bit or 64-bit values.
When working with 64-bit indexes, we require 64-bit integer divisions which are slower than 32-bit integer divisions on x64 processors. We always use the same 64-bit random-number generator, but in the 32-bit case, we only use the least significant 32~bits of each random word. 
For reference, in both the 32-bit and 64-bit figures, we include the results obtained with the shuffle functions of the standard library (\texttt{std::shuffle}), implemented using our 64-bit random-number generator. For small arrays, the \texttt{std::shuffle} has performance similar to our implementation of the OpenBSD algorithm, but it becomes slightly slower when shuffling larger arrays.

\begin{figure}\centering
\includegraphics[width=0.49\columnwidth]{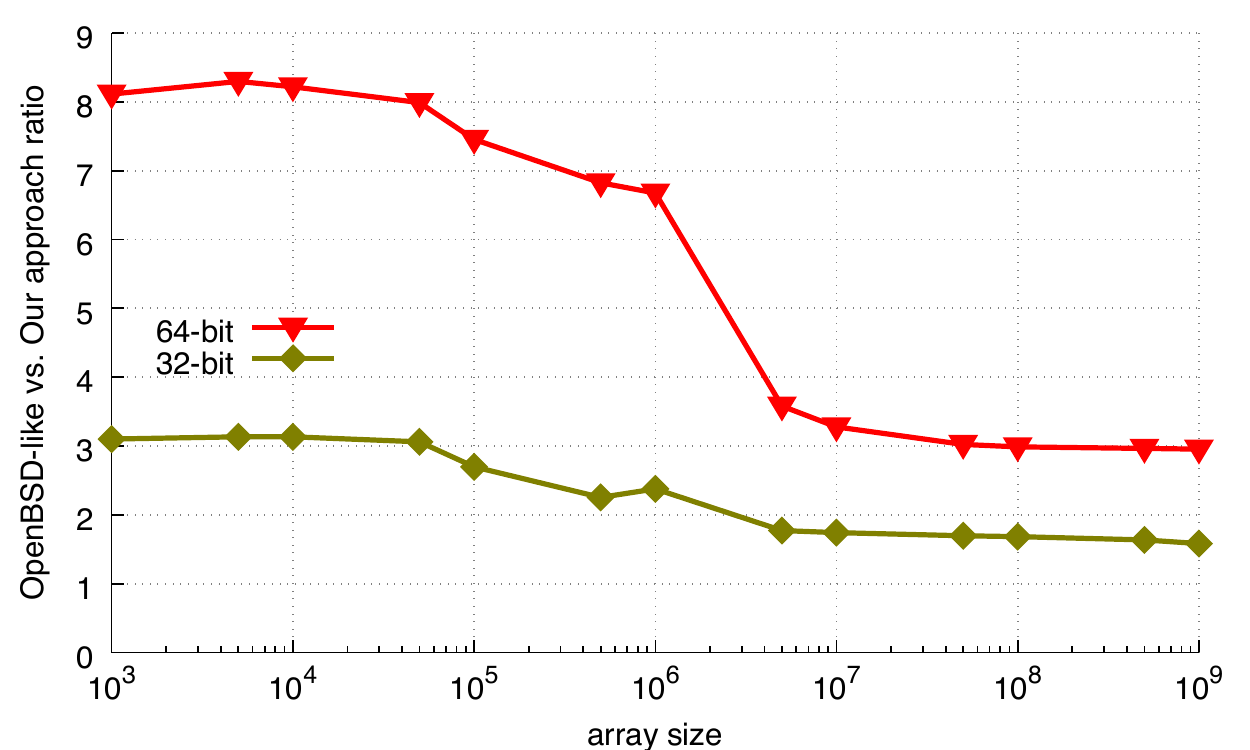}
\caption{\label{fig:shuffleratio}Ratio of the timings of the OpenBSD-like approach and of our approach.}
\end{figure}

\begin{figure}\centering
\subfloat[32-bit indexes \label{fig:indexes32}]{%
\includegraphics[width=0.49\columnwidth]{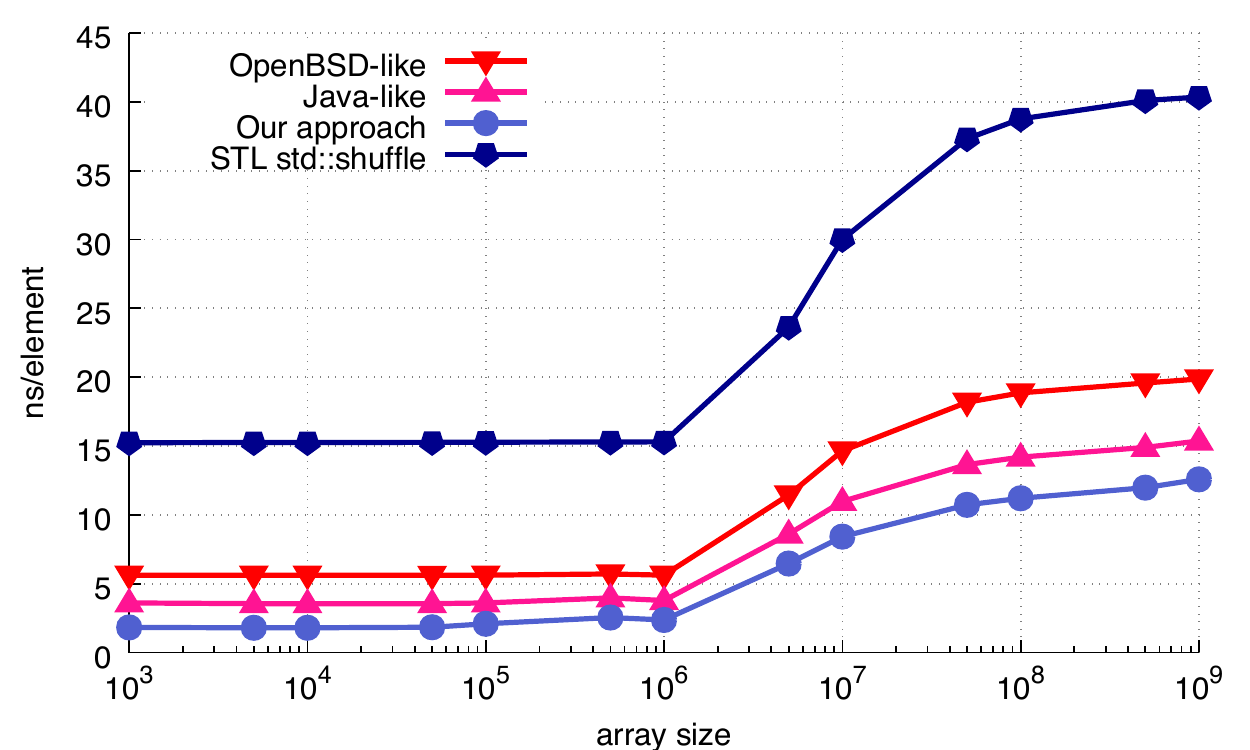}
}
\subfloat[64-bit indexes \label{fig:indexes64}]{%
\includegraphics[width=0.49\columnwidth]{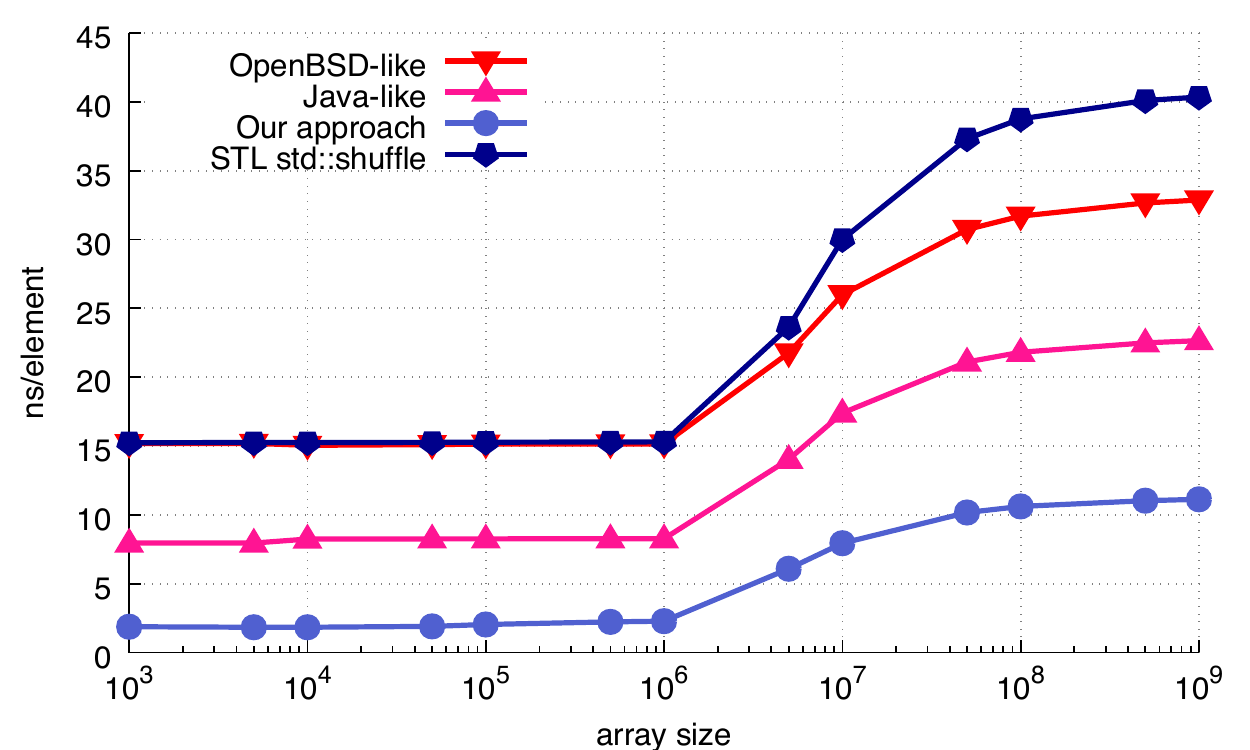}
}
\caption{\label{fig:fastshuffle}Wall-clock time in nanoseconds per element to shuffle arrays of random integers.}
\end{figure}

We present our experimental results in Fig.~\ref{fig:fastshuffle}. We report the wall-clock time averaged over at least 5~shuffles. When the arrays fit in the cache, we expect them to remain in the cache.  The time required is normalized by the number of elements in the array. As long as the arrays fit in the CPU cache, the array size does not affect the performance. As the arrays grow larger, the latency of memory access becomes a factor and the performance decreases. 
\begin{itemize}
\item In the 32-bit case, the approach with few divisions can  be almost twice as fast as the Java-like approach which  itself can  be at least 50\% faster than the OpenBSD-like approach. 
\item  When shuffling with 64-bit indexes as opposed to 32-bit indexes, our implementations of the  OpenBSD-like and Java-like algorithms become significantly slower (up to three times) due to the higher cost of the 64-bit division. 
    Thus our approach can be more than three times faster than the Java version in the 64-bit case.
\end{itemize}
The relative speed differences between the different algorithms become less significant when the arrays grow larger.  In Fig.~\ref{fig:shuffleratio}, we present the ratio of the OpenBSD-like approach with our approach. We see that the relative benefit of our approach diminishes when the array size increases. In the 32-bit case, for very large arrays, our approach is merely 50\% faster whereas it is nearly three times faster for small arrays. Using Linux \texttt{perf}, we estimated the  number of cache misses to shuffle an array containing 100~million integers and found that the OpenBSD approach generates about 50\% more cache misses than our approach.

%

\begin{figure}\centering
\subfloat[32-bit indexes \label{fig:indexes32precompopenbsd}]{%
\includegraphics[width=0.49\columnwidth]{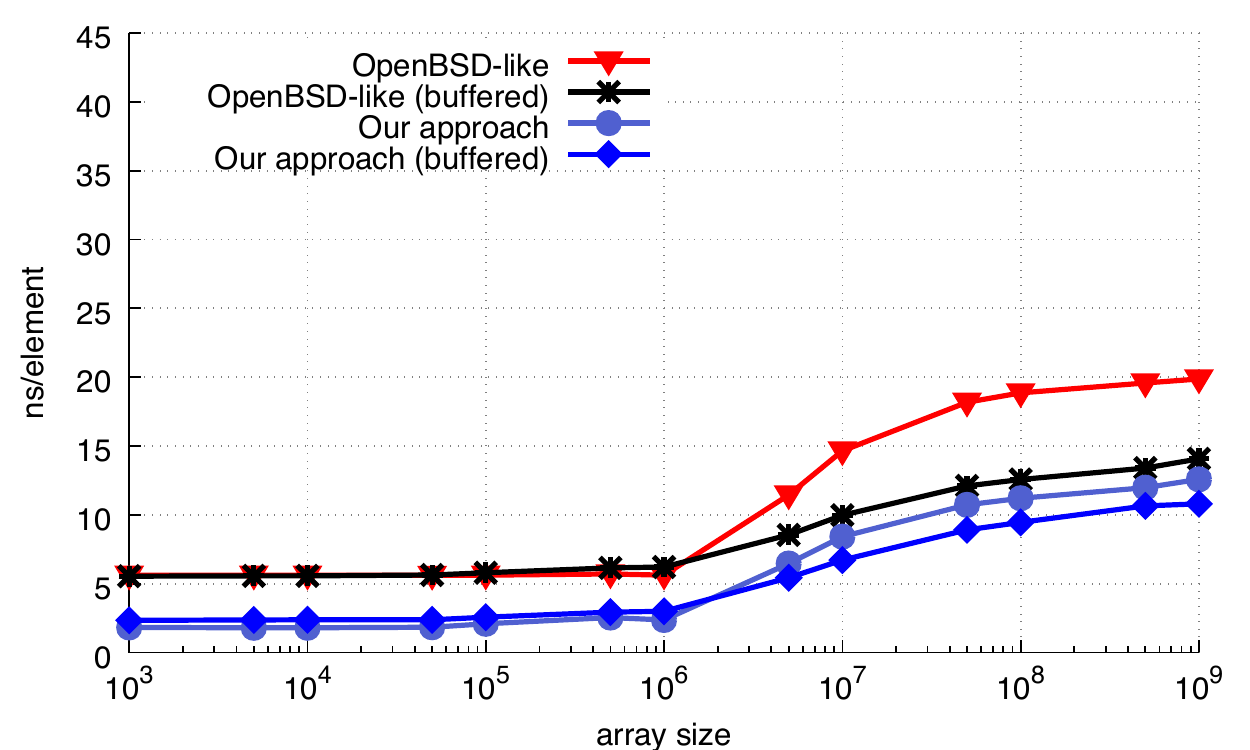}
}
\subfloat[64-bit indexes \label{fig:indexes64precompopenbsd}]{%
\includegraphics[width=0.49\columnwidth]{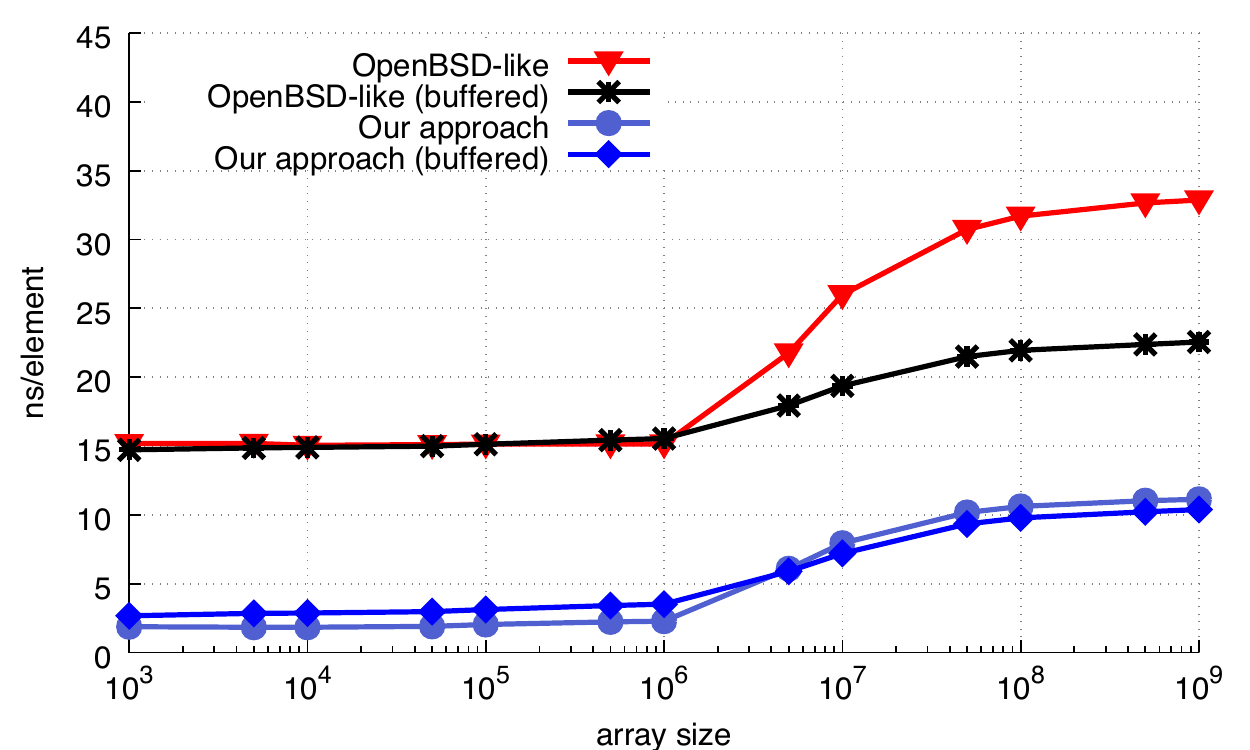}
}
\caption{\label{fig:bufferedfastshuffle}Wall-clock time in nanoseconds per element to shuffle arrays of random integers using either regular shuffles or buffered shuffles with a buffer of size 256 (see Algorithm~\ref{algo:bufferedknuthshuffle}).}
\end{figure}

To help the processor prefetch memory and reduce the number of
cache misses, we can compute the random integers in small blocks, and then shuffle while reading the precomputed integers (see Algorithm~\ref{algo:bufferedknuthshuffle}). The resulting buffered algorithm
is  equivalent to the conventional Fisher-Yates random shuffle, and it involves the computation of the same number of random indexes, but it differs on how the memory accesses are scheduled.
In Fig.~\ref{fig:bufferedfastshuffle}, we see that the OpenBSD-like approach benefits from the buffering when shuffling large arrays. A significant fraction of the running time of the regular OpenBSD-like implementation is due to caching issues.  When applied to our approach, the benefits of the buffering  are small, and for small to medium arrays, the buffering is slightly harmful.

\begin{algorithm}[t]
\begin{algorithmic}[1]
\REQUIRE array $A$ made of $n$ elements indexed from $0$ to $n-1$
\REQUIRE $B\leftarrow$ a small positive constant (the buffer size)
\STATE $i \leftarrow n-1$
\STATE $Z\leftarrow $ some array with capacity $B$ (the buffer)
\WHILE{$i\geq B$}
\FOR{$k \in \{i, i-1, \ldots, i -B + 1\}$}
\STATE $Z_k \leftarrow $random integer in $[0,k]$
\ENDFOR
\FOR{$k \in \{i, i-1, \ldots, i -B + 1\}$}
\STATE exchange $A[k]$ and $A[Z_k]$
\ENDFOR
\STATE $i \leftarrow i - B$
\ENDWHILE
\WHILE{$i>0$}
\STATE $j \leftarrow $ random integer in $[0,i]$
\STATE exchange $A[i]$ and $A[j]$
\STATE $i \leftarrow i - 1$
\ENDWHILE
\end{algorithmic}
\caption{Buffered version of the Fisher-Yates random shuffle\label{algo:bufferedknuthshuffle}.}
\end{algorithm}

\begin{figure}\centering
\subfloat[32-bit indexes \label{fig:floatindexes32}]{%
\includegraphics[width=0.49\columnwidth]{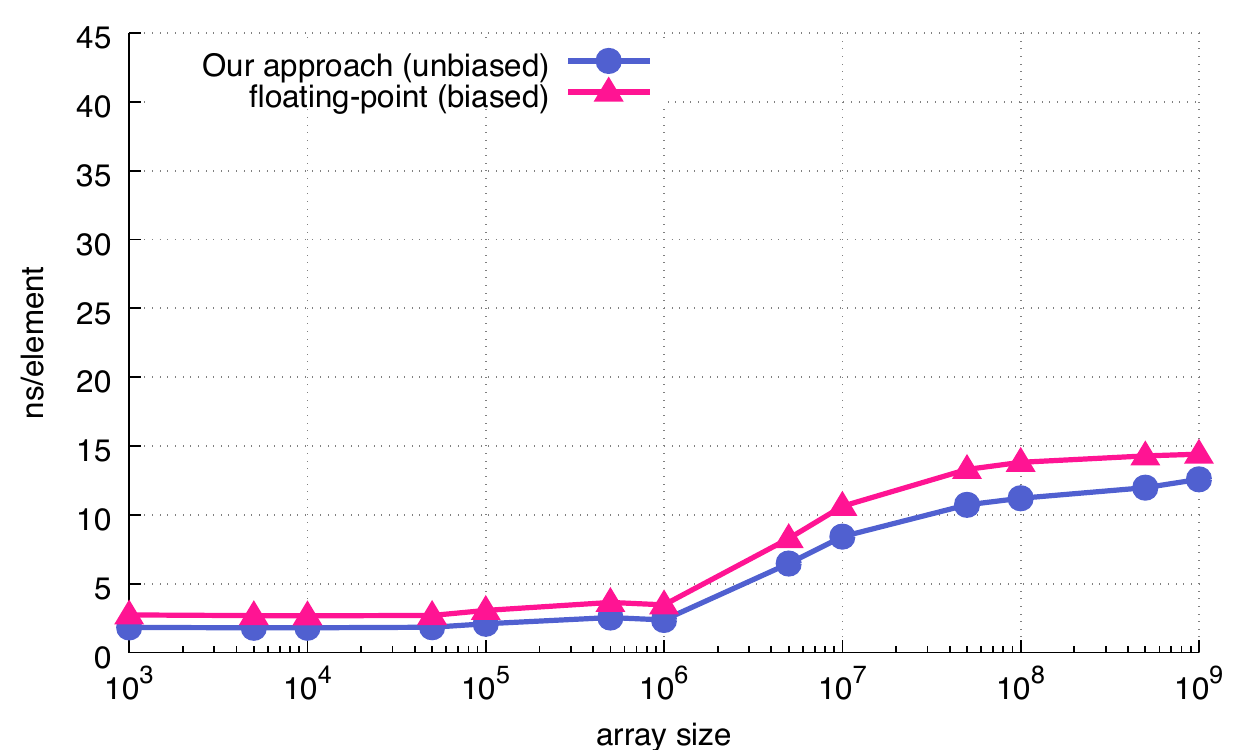}
}
\subfloat[64-bit indexes \label{fig:floatindexes64}]{%
\includegraphics[width=0.49\columnwidth]{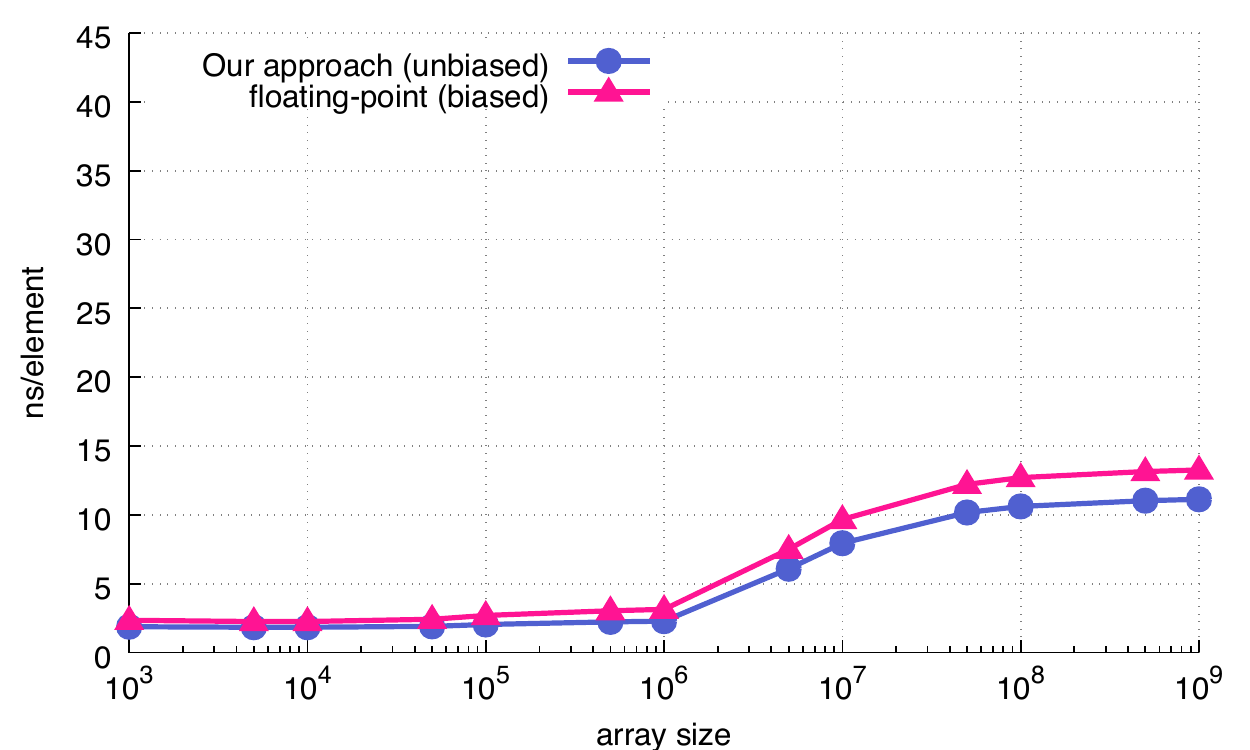}
}
\caption{\label{fig:floatfastshuffle}Wall-clock time in nanoseconds per element to shuffle arrays of random integers using either Algorithm~\ref{algo:divisionless} or an approach based on floating-point multiplications.}

\end{figure}

We also compare  Algorithm~\ref{algo:divisionless}  to the floating-point approach we described in \S~\ref{sec:intro}
where we represent the random number as a floating-point value in $[0,1)$ which we multiply by $s$ to get a number in $[0,s)$. As 
illustrated in Fig.~\ref{fig:floatfastshuffle}, the floating-point
approach is slightly slower (10\%--30\%) whether we use 32-bit or 64-bit indexes. Moreover, it introduces a small bias and it is limited to $s\leq 2^{24}$ in the 32-bit case and to $s\leq 2^{53}$ in the 64-bit case.

\section{Conclusion}

We find that the algorithm often used in OpenBSD and macOS (through the \texttt{arc4random\_uniform} function) requires two divisions
per random number generation.
It is also the slowest in our tests.
The Java approach that often requires  only one division,
can be faster. We believe that it should be preferred to the 
 OpenBSD algorithm.

As we have demonstrated, we can use nearly no division
at all and at most one in the worst case. Avoiding divisions can multiply the speed of unbiased random shuffling functions on x64 processors.

For its new random shuffle function, the Go programming language adopted our proposed approach~\cite{Snyder2017}. The Go authors justify this decision by the fact that it results in 30\% better speed compared with the application of the OpenBSD approach (Algorithm~\ref{algo:arc4}).

 Our results are only relevant in the context where the generation
of random integers is fast compared with the latency of a division operation. 
In the case where the generation of random bits is likely the bottleneck, other approaches would be preferable~\cite{Bacher:2017:GRP:3040971.3009909}. 
Moreover, our approach may not be applicable to specialized processors such as  Graphics Processing Units (GPUs) that lack support for the computation of the full multiplication~\cite{Langdon:2009:FHQ:1570256.1570353}.\footnote{The CUDA API  provided by  Nvidia offers relevant intrinsic functions such as \texttt{\_\_umul64hi}.}

\appendix
\section{Code Samples}
\label{appendix:codesamples}
\begin{minipage}{\linewidth}\begin{lstlisting}
// returns value in [0,s)
// random64 is a function returning random 64-bit words
uint64_t openbsd(uint64_t s, uint64_t (*random64)(void)) {
  uint64_t t = (-s) % s;
  uint64_t x;
  do {
    x = random64();
  } while (x < t);
  return x % s;
}\end{lstlisting}\end{minipage}
\begin{minipage}{\linewidth}\begin{lstlisting}
uint64_t java(uint64_t s, uint64_t (*random64)(void)) {
  uint64_t x = random64();
  uint64_t r = x % s;
  while (x - r > UINT64_MAX - s + 1) { 
    x = random64();
    r = x % s;
  }
  return r;
}\end{lstlisting}\end{minipage}
\begin{minipage}{\linewidth}
\begin{lstlisting}
uint64_t nearlydivisionless(uint64_t s, uint64_t (*random64)(void)) {
  uint64_t x = random64();
  __uint128_t m = (__uint128_t) x * (__uint128_t) s;
  uint64_t l = (uint64_t) m;
  if (l < s) {
   uint64_t t = -s % s;
    while (l < t) {
      x = random64();
      m = (__uint128_t) x * (__uint128_t) s;
      l = (uint64_t) m;
    }
  }
  return m >> 64; 
}\end{lstlisting}
\end{minipage}

\begin{acks}
The work is supported by the \grantsponsor{NSERC}{Natural Sciences and Engineering Research Council of Canada}{} 
under grant
 number~\grantnum{NSERC}{RGPIN-2017-03910}. The author would like to thank R.~Startin and J.~Epler for independently reproducing the experimental results and providing valuable feedback.
\end{acks}

\bibliographystyle{ACM-Reference-Format}

\bibliography{rangedrng}
\end{document}